\documentclass[11pt,letter]{article}
\usepackage{graphicx} % Required for inserting images
\usepackage{fontawesome5}
\usepackage{hyperref}

\usepackage[utf8]{inputenc}
\usepackage[english]{babel}
\usepackage{amsthm}
\usepackage{amsmath}
\usepackage{comment}
\usepackage[ruled,vlined,linesnumbered]{algorithm2e}
\usepackage[table]{xcolor}
\usepackage{amsfonts}

\newcommand{\prev}{\mathsf{prev}}
\newcommand{\lprev}{\mathsf{lprev}}
\newcommand{\Alive}{\mathsf{Alive}}

\newcommand{\Oalive}{\mathcal{O}_{\mathrm{alive}}}
\newcommand{\Owit}{\mathcal{O}_{\mathrm{wit}}}
\newcommand{\OlprevT}{\mathcal{O}_{\mathrm{lprev}}^{T}}
\newcommand{\Osample}{\mathcal{O}_{\mathrm{sample}}}
\newcommand{\Omis}{\mathcal{O}_{\mathrm{mis}}}
\newcommand{\Overify}{\mathcal{O}_{\mathrm{verify}}}
\newcommand{\Oblock}{\mathcal{O}_{\mathrm{block}}}
\newcommand{\Oplprev}{\mathcal{O}_{\mathrm{lprev}}^{P}}
\newcommand{\Otlprev}{\mathcal{O}_{\mathrm{lprev}}^{T}}
\makeatletter
\def\cramped        % "Cramped" list style; put after \begin{itemize}
{\parskip\@outerparskip\@topsep3pt plus 1pt minus 1pt
  \@topsepadd3pt plus 1pt minus 1pt\itemsep1pt             
}
\makeatother

\def\paragraph#1{\par\smallskip\noindent{\bf #1\ \ }%
\ignorespaces
}

\newbox\tallstrutbox
\setbox\tallstrutbox=\hbox{\vrule height8pt depth 3.5pt width0pt}
\def\tallstrut{\relax\ifmmode\copy\tallstrutbox\else\unhcopy\tallstrutbox\fi}
\newbox\tallerstrutbox
\setbox\tallerstrutbox=\hbox{\vrule height10pt depth 3.5pt width0pt}
\def\tallerstrut{\relax\ifmmode\copy\tallerstrutbox\else\unhcopy\tallerstrutbox\fi}

\def\path{\mathsf{path}}

\long\def\symbolfootnote[#1]#2{\begingroup%
\def\thefootnote{\fnsymbol{footnote}}\footnote[#1]{#2}\endgroup}

\newtheorem{corollary}{Corollary}
\newtheorem{lemma}{Lemma}

\newtheorem{theorem}{Theorem}

\newtheorem{definition}{Definition}
\usepackage{colortbl} 

\title{Quantum Algorithm for Parameterized Aperiodic Pattern Matching}

\author{Rahul Shah \textsuperscript{\href{mailto:rahul@lsu.edu}{\faEnvelope}} \\ Lalita Sthapit \textsuperscript{\href{mailto:lsthap1@lsu.edu}{\faEnvelope}} \\ Division of Computer Science and Engineering \\ Louisiana State University \\ [0.3em]}

\begin{document}

\maketitle

\begin{abstract}   

In string matching, the most basic setting is that we are given a text $T[0 \ldots n-1]$ and a pattern $P[0\ldots m-1]$. We want to find whether $P$ appears in $T$ as a substring or not. This is a very fundamental problem with wide applicability across many practical domains. We consider a variant of string matching namely, parameterized pattern matching, in which, two strings are considered to match if there exists a one-to-one mapping between the symbols appearing in them. Thus, two strings can match even if they consist of different alphabet symbols, provided that the pattern of symbols repetitions is identical in both strings.

Bringing out quantum advantage in string matching problems involves a careful blend of string combinatorics and quantum primitives. In the RAM model, the best time complexity for pattern matching is $O(n+m)$ or simply written as $O(n)$, as $m$ is taken to be $\le n$. In the quantum query model, Ramesh and Vinay ~\cite{RameshVinay2003} established a $\widetilde{O}(\sqrt{n}+\sqrt{m})$ query complexity for the conventional string matching problem. In this work, we study quantum query complexity of parameterized pattern matching focusing on non-periodic patterns and present a quantum algorithm with $\widetilde{O}(\sqrt{n}+\sqrt{m})$ query complexity, assuming oracle access to the input text and pattern, as well as their precomputed previous-encoding arrays.

\end{abstract}

\section{Introduction}
String matching is one of the most fundamental problems in computer science. Given a $T[0  ~\ldots~  n-1]$ and a pattern $P[0  ~\ldots~  m-1]$, the goal is to find whether $P$ occurs in $T$ as a contiguous sub-string. This problem appears in many areas, including text processing, data compression, biological sequence analysis, information retrieval, version control, and program analysis.

In the classical RAM model, exact string matching can be solved in linear time. In the quantum query model, however, one may hope for sublinear algorithms. One of the central results in this direction is due to Ramesh and Vinay~\cite{RameshVinay2003}, who presented that exact string matching can be solved in  $\widetilde{O}(\sqrt{n} + \sqrt{m})$ quantum time. Their algorithm combines Grover's search ~\cite{Grover1996} with Vishkin's deterministic sampling technique~\cite{Vishkin1991}. The deterministic sample reduces the number of candidate alignments that need to be verified, and Grover search is then used to search over the remaining candidates efficiently.

In this paper, we study quantum algorithms for parameterized pattern matching for an aperiodic pattern. Our main objective is to understand whether deterministic-sampling-style candidate reduction can be extended beyond ordinary exact string matching and combined with quantum search in this matching models.

Parameterized pattern matching was introduced by Baker~\cite{Baker1996}, motivated by application such as detecting duplicated code where variable names may change but the structural pattern remains the same. In parameterized pattern matching, the pattern and the text sub-string do need to be equal character-by-character. Instead, they match if there is a one-to-one renaming of symbols that makes them equal. Two strings parameterize-match if and only if their local previous-encoding are equal. Thus, parameterized matching compares the recurrence structure of symbols rather than the symbols themselves.

This creates the difficulty in the quantum setting. For a fixed alignment $a$, the local previous-encoding value of a text position depends on whether the previous occurrence of the same symbol lies inside the current text window $T[a  ~\ldots~  a+m-1]$. Therefore, the value observed at a text position depends on the left boundary of an alignment. A direct quantum approach is possible. For each text alignment $a$, we can use Grover search over the $m$ pattern positions to find a position where the previous-encoding values disagree. This verifies one alignment in $O(\sqrt{m})$ time. Then we can use another Grover search over the $O(n)$ possible alignments. This gives total complexity of $O(\sqrt{n}\sqrt{m})=O(\sqrt{nm})$. This is quadratic improvement over the naive $O(nm)$ approach, but it is still weaker than $\widetilde{O}(\sqrt{n} + \sqrt{m})$ bound known for exact string matching.

This gap motivates the main question of this work: can deterministic sampling be extended to parameterized matching in a way that reduces the naive matching complexity? This extension is not immediate. In exact matching, a mismatch position can be found by directly comparing the symbols at the corresponding positions. In parameterized matching, identifying such mismatch is more complicated because the previous-encoding values change depending on the starting position of the text window being considered.

\subsection{Related Work}
Classical exact string matching has a long history. Knuth, Morris and Pratt~\cite{Knuth1977} introduced an algorithm based on failure function, while Boyer and Moore ~\cite{Boyer1977} developed another efficient algorithm that reduces unnecessary character comparisons by using suitable shifts. These algorithms solve the exact string matching problem in linear-time. Vishkin~\cite{Vishkin1991} introduced deterministic sampling as a powerful combinatorial technique to reduce candidate alignments in pattern matching. For an aperiodic pattern, logarithmic deterministic sample eliminates all but one candidate alignment within a local neighborhood. Extending deterministic sampling technique to parametric strings is not straightforward since they exhibit different structural properties. Apostolico et.al ~\cite{ApostolicoG08} studied periodicity in parameterized strings and showed that classical periodicity results, including Fine-Wilf Lemma, do not directly extend to this setting. In a different direction Jalsenius et.al ~\cite{JalseniusPS13} studied parameterized matching in the streaming model and developed an algorithm using sublinear space.

Baker~\cite{Baker1996} introduced parameterized suffix trees and introduced the previous-encoding representation, which provides a way to check whether two strings parameterize-match. Amir et al~\cite{Amir1994} studied the alphabet dependence of parameterized matching and presented an efficient algorithm for finding all parameterized occurrences of a pattern in a text. Further research has explored extensions to multiple-pattern and approximate parameterized matching~\cite{Idury1996,Hazay2007}, as well as the reconstruction of parameterized strings from suffix array representation~\cite{Amir2022}. These works highlight the continued development of parameterized matching and its related problems. 

Quantum search algorithms provide an important tools for designing faster algorithms for search problems. Grover~\cite{Grover1996} introduced a quantum algorithm for searching an unsorted database of size $N$ using $O(\sqrt{N})$ queries, giving a quadratic improvement over classical search. Boyer~\cite{Boyer1996}, Brassard, H{\o}yer and Tapp~\cite{BrassardHoyerTapp1998} studied tight bounds for quantum searching, including settings where the number of marked items is unknown. D\"urr and H{\o}yer~\cite{DurrHoyer1996} gave a quantum algorithm for minimum finding, which is also very useful in quantum string algorithms. These primitives are used in the quantum pattern matching framework to search over text blocks, candidate alignments, and witness positions. Quantum algorithms for string problems beyond exact matching have also been studied. Niroula and Nam~\cite{Niroula2021} studied quantum circuit implementations of string matching, while Wang and Ying~\cite{wang2024quantum} and Akmal and Jin~\cite{Akmal2023} developed quantum algorithms for problems such as lexicographically minimal string rotation, and longest common substring. More recently, Gibney et al~\cite{Gibney2024} studied quantum algorithms for bounded edit distance and Lempel-Ziv factorization. These indicate the growing use of quantum techniques in string problems.

In this paper we present a quantum algorithm for parameterized pattern matching for p-aperiodic patterns. Our matching algorithm follows the basic framework of Ramesh and Vinay~\cite{RameshVinay2003}. However, extending this approach to parameterized setting introduces additional challenges. For classical string, deterministic sampling and pattern matching relies on comparing corresponding characters in shifted copies of the string and pattern. In parameterized string, such comparison must consider local previous-encoding, whose values changes depending on the starting position of the substring and text-window being examined. Thus, both the construction of deterministic sample and subsequent matching process must keep account for these changing left boundaries and their corresponding local previous encodings. To address this, we define deterministic-sample, ricochet property and quantum oracles for the parameterized setting. We construct a quantum friendly randomized sample of size $O(\log m)$ and use it for candidate-filtering. Assuming oracle access to the inputs as well as to precomputed previous-encoding array, our pattern matching algorithm achieves quantum query complexity $\widetilde{O}(\sqrt{n} + \sqrt{m})$ for p-aperiodic pattern.

\section{Preliminaries}

Let $T[0 ~\ldots~ n-1]$ be a string of $n$ characters, each drawn from 
an set of alphabet symbols $\Sigma$ of size $\sigma$. For convenience, we may assume $\Sigma = \{0 ~\ldots~ \sigma-1\}$.

In pattern matching problem, we are given a pattern $P[0 ~\ldots~ m-1]$ whose characters are also drawn from $\Sigma$. Central question in the field of pattern matching is: does $P$ exist as a substring of $T$? More formally, is there a location $i$ such that $T[i ~\ldots~ i+m-1] = P[0 ~\ldots~ m-1]$?

In parameterized matching problem, $T$ is drawn from alphabet set $\Sigma$ and $P$ may be drawn from a different alphabet set $\Sigma'$. A pattern $P[0 ~\ldots~ m-1]$ matches text substring $T[i ~\ldots~ i+m-1]$ if there is a bijection $\pi: \Sigma \rightarrow \Sigma'$ such that $P[j] = \pi(T[i+j])$ for all $j \in [0 ~\ldots~ m-1]$.

\subsection{Previous Encoding}
\begin{definition}
Given a string $S[0 ~\ldots~ s-1]$ it's Prev (or Previous) encoding $\prev_S$ is given by:
\begin{itemize}
\cramped
    \item $\prev_S[i] = 0$ if $S[i] \ne S[j] ~~\forall j < i$.
    \item $\prev_S[i] = i-k$ where $k < i$ is a greatest index such that $T[k] = T[i]$
\end{itemize}
\end{definition}

$\prev_S[i]$ records the distance between $i$ and the previous occurrence of symbol $T[i]$ in $T$. Prev encoded strings have the alphabet set of $[0 ~\ldots~ s-1]$.

Let's denote suffix $S[l ~\ldots~  m-1]$ as string $S_l$.
\begin{definition}
Local Prev Encoding with respect to starting index $l$ is denoted by $\lprev_S^l$. For any index $i \ge l$, $\lprev^l_S[i] = \prev_{S_l}[i-l]$.

\end{definition}

Given an array $\prev_S$ we can compute $\lprev^k_S[i] = 0$ if $\prev_S[i] > i - k$ and $\lprev^k_S[i] = \prev_S[i]$ otherwise. 

\begin{center}
\begin{tabular}{ccccccccc}
     $i$ & 0 & 1 & 2 & 3 & 4 & 5 & 6 & 7 \\ \hline
     $S[i]$ & a & b & a & b & c & a & b & c \\ 
     $\prev_S[i]$ & 0 & 0 & 2 & 2 & 0 & 3 & 3 & 3 \\ 
     $\lprev_S^3[i]$ & -- & -- & -- & 0 & 0 & 0 & 3 & 3 \\
     \hline
\end{tabular}
\end{center}

\begin{lemma}\label{prev} \cite{Baker1996}
Two strings $S[0 ~\ldots~ m-1]$ and $T[0 ~\ldots~  m-1]$ are a parameterized match if and only if $\prev_S = \prev_T$.
\end{lemma}

This means for all $i \in [0 ~\ldots~  m-1]$, $\prev_S[i] = \prev_T[i]$.

\begin{lemma}
\label{lem:computation-in-O-1}
Given precomputed array $\prev_S[\cdot]$, $\lprev$ for any particular index can be computed in $O(1)$ time.
\end{lemma}
\begin{proof}
     By Definition 3, the local previous encoding depends on whether the previous occurrence of \(S[i]\) lies within the suffix \(S[k\ldots]\). If \(\prev_S[i]>i-k\), the previous occurrence lies before \(k\), and hence \(\lprev_S^k[i]=0\). Otherwise, the previous occurrence lies within the suffix, and \(\lprev_S^k[i]=\prev_S[i]\). Thus,

\[
\lprev_S^k[i]=
\begin{cases}
0 & \text{if } \prev_S[i]>i-k,\\
\prev_S[i] & \text{otherwise}.
\end{cases}
\]

Since \(\prev_S\) is precomputed, the value can be obtained using a constant number of operations. Therefore, the query takes \(O(1)\) time. 
\end{proof}
 
\begin{definition}
    
A string $P[0 ~\ldots~  m-1]$ is called p-periodic with period $s \le m/2$ if $P[0 ~\ldots~  m-s-1]$ and $P[s ~\ldots~  m-1]$ are a parameterized match.
If no such $s$ exists, then the pattern is p-aperiodic. 
\end{definition}

\subsection{Deterministic Sample}
Let $P[0\ldots m-1]$ be a p-aperiodic pattern. A distinguishing shift is an integer $k \leq m/2$ for which there exists a set of positions $\delta_1 <\delta_2 <  ~\ldots~   < \delta_d$ satisfying $k\leq \delta_i\leq k + m/2$ such that for each $i \in \{1,..d\}$ and for all $j \in [0,m/2)$ with $j \neq k$, $
\lprev^{k}_{P}[\delta_i]
\neq
\lprev^{j}_{P}[\delta_i-k+j]$.
Then, the tuple $(k; \delta_1, \delta_2,  ~\ldots~  ,\delta_d)$ is called a deterministic sample of $P$. Similar to \cite{Vishkin1991} we can show that in parameterized case that:

\begin{lemma}\cite{Vishkin1991}
    The deterministic sample of $P[0 ~\ldots~  m-1]$ consists of $d = O(\log m)$ sample indices.
\end{lemma}

If a pattern is periodic with minimum period $s$, then deterministic sample is a collection $K$ of all $k + as \le m/2$ ($a = 0, 1,  ~\ldots~   $).
%along with indices $\delta_1+ as, \delta_2 + as,  ~\ldots~   , \delta_d + as$. 

\subsection{Oracle and Quantum Search}
\paragraph{Oracle}
In quantum model, an oracle is a black box unitary operation that provides access to a function $f:X \rightarrow Y$. Each application of an oracle is counted as one quantum query. The complexity of an algorithm is the number of oracle queries it makes. In this paper, we use oracles to access information about the pattern, text and their previous-encodings, as well as to perform other operations required by the algorithm. The formal definitions of these problem-specific oracles are provided in Appendix section.\\

\paragraph{Grover's Search Algorithm ~\cite{Grover1996}}
Let $X=\{0,1, \ldots,N-1\}$ be an unstructured search space, and let $f:X\rightarrow\{0,1\}$ be a Boolean Oracle, where $f(x)=1$ indicates a marked element. The objective is to find out an $x \in X$ such that $f(x)=1$. If at least one marked element exists, Grover's algorithm finds a marked element with bounded error using $O(\sqrt{N})$ quantum oracle queries.\\

\paragraph{Durr-Hoyer Minimum Finding Algorithm~\cite{DurrHoyer1996}}
Let $X=\{0,1,\ldots,N-1\}$ be a set of $N$ elements, and let $g:X \rightarrow \mathbb{R}$ be an evaluation function. The objective is to find
\[
x_{min} = \operatorname*{arg\,min}_{x\in X} g(x).
\]
The quantum minimum-finding algorithm of Durr-Hoyer identifies $x_{min}$ with bounded error using $O(\sqrt{N})$ quantum queries to the evaluation oracle. By replacing the minimum operation with a maximum operation, the same approach can be used to find 
\[
x_{max} = \operatorname*{arg\,max}_{x\in X} g(x).
\]
with $O(\sqrt{N})$ quantum queries.

%----------_____________---------------______
\section{Parameterized Pattern Matching}

\subsection{Parameterized Periodicity and Witness}
Let $P[0..m-1]$ be a pattern of length $m$. In parameterized matching, exact character equality is replaced by parameterized equivalence. Let us use the notation $X \approx_p Y$ to denote that two strings $X$ and $Y$ parameterize-match with each other. Equivalently, $X \approx_p Y$ if and only if their local previous-encodings are equal, that is, $ \prev_X = \prev_Y$. 
In this work, we assume that the previous-encoding arrays $\prev_P$ and $\prev_T$ are precomputed and available for constant-time oracle access. 

\begin{definition}[Parameterized period]
A shift $s$, where $1 \leq s < m$, is called a parameterized period of $P[0 ~\ldots~  m-1]$ if
\[
P[0 ~\ldots~  m-s-1] \approx_p P[s ~\ldots~  m-1]
\]
Equivalently, $s$ is a parameterized period if $\prev(P[0 ~\ldots~  m-s-1]) = \prev(P[s ~\ldots~  m-1])$. The pattern $P$ is called parameterized-periodic, or p-periodic, if it has a parameterized period $s$ satisfying $1 \leq s \leq \lfloor \frac{m}{2} \rfloor$.
Otherwise, $P$ is called parameterized-aperiodic, or p-aperiodic. Thus, $P$ is $p-aperiodic$ if for every shift $s$, $
1 \leq s \leq \lfloor \frac{m}{2} \rfloor,
$
we have $
\prev(P[0 ~\ldots~  m-s-1]) \neq \prev(P[s ~\ldots~  m-1]). $
\end{definition}

We define parameterized witnesses relative to an unshifted copy of $P$ starting at position $0$. If the two overlapping copies start at position $a$ and $a+s$, respectively, then a witness position $t$ corresponds to the absolute position $a+t$.
\begin{definition}[Parameterized witness]
Let $P[0 ~\ldots~  m-1]$ be p-aperiodic and let $ 1 \leq s \leq \lfloor \frac{m}{2} \rfloor$. A position $t$, where $0 \leq t < m-s$, is called a parameterized witness for a shift $s$ if $\prev(P[0 ~\ldots~  m-s-1])[t] \neq \prev(P[s ~\ldots~  m-1])[t]$.
\end{definition}

\begin{lemma}[Existence of parameterized witness]
Let $P[0 ~\ldots~  m-1]$ be p-aperiodic. Then for every shift $1 \leq s \leq \lfloor \frac{m}{2} \rfloor$, there exists at least one parameterized witness.
\end{lemma}
\begin{proof}
    Since $P$ is p-aperiodic, shift $s$ is not a parameterized period. Therefore, $
    P[0 ~\ldots~  m-s-1] \not\approx_p P[s ~\ldots~  m-1]$.
    This implies $\prev(P[0 ~\ldots~  m-s-1]) \neq \prev(P[s ~\ldots~  m-1])$. Since these two previous-encoding strings are not equal, there exists at least one position $t$, with $0 \leq t < m-s$, such that $
    \prev(P[0 ~\ldots~  m-s-1])[t] \neq \prev(P[s ~\ldots~  m-1])[t]$.
    By definition, such a position $t$ is a parameterized witness for the shift $s$.
\end{proof}
    
\subsection{Ricochet Property for Parameterized Matching}
P-aperiodic patterns satisfy a Ricochet property analogous to the classical aperiodic case. Two occurrences of a p-aperiodic pattern cannot start too close to each other because if they did, their overlap would force a parameterized period.

Let $T[0 ~\ldots~  n-1]$ be a text and $P[0 ~\ldots~  m-1]$ be a pattern. We say that $P$ has a parameterized occurrence at position $i$ in $T$ if $
P \approx_p T[i ~\ldots~  i+m-1]. $

\begin{lemma}[Overlapping occurrences imply a parameterized period]
\label{lem:overlap-occ-imply-pperiod}
Suppose $P[0 ~\ldots~  m-1]$ has two parameterized occurrences in $T$, starting at positions $i$ and $j$. Let $d=j-i$. If $d<m$, then $d$ is a parameterized-period of $P$. That is, $
P[0 ~\ldots~  m-d-1] \approx_p P[d ~\ldots~  m-1].$
Equivalently, $\prev(P[0 ~\ldots~  m-d-1]) = \prev(P[d ~\ldots~  m-1])$
\end{lemma}
\begin{proof}
Since $P$ occurs at position $i$, we have $P \approx_p T[i ~\ldots~  i+m-1].$
Since $P$ also occurs at position $j=i+d$, we have $
P \approx_p T[i+d  ~\ldots~  i+d+m-1].$
Because $d<m$, the two occurrences overlap. Their common text region is $
R=T[i+d ~\ldots~  i+m-1]$
which has length $m-d$.
In the first occurrence, this shared text region corresponds to the suffix $P[d ~\ldots~  m-1]$. In the second occurrence, the same shared text region corresponds to the prefix $P[0 ~\ldots~  m-d-1]$. Therefore, $P[d ~\ldots~  m-1] \approx_p R$ and $P[0 ~\ldots~  m-d-1] \approx_p R$. By transitivity, $
P[0 ~\ldots~  m-d-1] \approx_p P[d ~\ldots~  m-1].$
Thus, $d$ is a parameterized-period of $P$.
\end{proof}

\begin{corollary}[Ricochet property of p-aperiodic patterns]
\label{cor:p-Rico}
Let $P[0 ~\ldots~  m-1]$ be p-aperiodic. Then no two distinct parameterized occurrences of $P$ in $T$ can start within distance $m/2$ of each other. That is, if $i$ and $j$ are two distinct occurrence positions of $P$, then $
|i-j| > \lfloor \frac{m}{2} \rfloor.$
\end{corollary}

\begin{proof}
Let us assume, for contradiction, that two distinct parameterized occurrences of $P$ start at positions $i$ and $j$ and satisfy $
|i-j| \leq \lfloor \frac{m}{2} \rfloor.$
Without loss of generality, assume $i<j$, and let $d=j-i$. Then $
1 \leq d \leq \lfloor \frac{m}{2} \rfloor.$
Since $d<m$, the two occurrences overlap. By lemma~\ref{lem:overlap-occ-imply-pperiod}, $d$ is a parameterized period of $P$. Therefore, $P$ has a parameterized-period that is satisfying $
 d \leq \lfloor \frac{m}{2}\rfloor$.
This contradicts the assumption that $P$ is p-aperiodic. Hence, no two distinct parameterized occurrences of a p-aperiodic pattern can start within the distance $m/2$ of each other.
\end{proof}

\begin{lemma}[Sample Exclusion]
\label{lem:sample exclusion property}
Suppose $P$ has a true occurrence at position $j$ in $T$. Then every alignment \(j'\) satisfying
$j' \in [j-s,\;j+\lfloor m/2 \rfloor - s]$ with $j' \neq j$ fails the deterministic sample test.
\end{lemma}

\begin{proof}
    Let $j' \neq j$ be an alignment in $[j-s,\;j+\lfloor m/2 \rfloor - s]$. By definition of the deterministic sample, there exists a sampled position that distinguishes a true occurrence at $j$ from the candidate alignment at $j'$. Their $\lprev$ values differ at that position. So, $j'$ fails the deterministic sample test.
\end{proof}

\begin{lemma}[Contrapositive of Ricochet property]
Let $I_b$ be a block, in the text, of size $m/2$. If $I_b$ contains more than one sample-passing alignments, then it cannot contain a true occurrence of $P$.
\end{lemma}
\begin{proof}
    Suppose, for contradiction, that $I_b$ contains a true occurrence beginning at position $j$ and also contains a second sample-passing alignment at a distinct position $j'$.
    Since, $|I_b|=m/2$, the two positions fall within the field of fire. This contradicts the fact that $j$ excludes other sample-passing alignments within the range $m/2$.
    Therefore, a block containing more than one sample-passing alignments cannot contain a true occurrence.
\end{proof}

\subsection{Algorithm for Parameterized Deterministic Sampling}
In this section, we describe deterministic sampling for parameterized pattern matching. The goal is to construct a small set of checkpoint positions in the pattern such that, inside any window of length roughly \(m/2\), at most one candidate occurrence can survive these checkpoint tests.

\paragraph{Checkpoints.}Initially, we consider all the possible shifts inside a window of length \(m/2\), $A_0 = \{0,1,\ldots,\lfloor m/2 \rfloor-1\}$. The set \(A_r\) denotes the set of shifts that survive after \(r\) checkpoint tests. At some round, let $p=\min A_r$ and $q=\max A_r$, and define $s=q-p$. Since \(P\) is p-aperiodic, \(s\) is not a parameterized period. Therefore, there exists a witness position \(t\), where \(0 \leq t < m-s\), such that $prev(P[0 ~\ldots~  m-s-1])[t] \neq prev(P[s ~\ldots~  m-1])[t]$. Lets convert this local witness position into a global checkpoint column by setting $i=q+t$. The common overlap interval of the shifted copies p and q is $L=q$ and $R=p+m-1$. For any surviving shift \(\gamma \in A_r\), we define its checkpoint value at column $i$ as $V(\gamma,i)$ = ${\lprev}^{L-\gamma}_P[i-\gamma]$. Since, $i -\gamma$ is interpreted relative to the shifted copy beginning at $\gamma$, $L-\gamma$ represents the start of the common overlap inside that shifted copy. 
%-------------------------------------------------------------

\begin{lemma}
    \label{lem:witness-separates}
    Let \(p=\min A_r\), \(q=\max A_r\), \(s=q-p\), and let \(t\) be a parameterized witness for shift \(s\). If \(i=q+t\) and \(L=q\), then $V(p,i)\neq V(q,i)$.
\end{lemma}
    
\begin{proof}
For the shifted copy beginning at $p$, the checkpoint column \(i=q+t\) corresponds to position $i-p = q+t-p = s+t$ inside the copy of $P$. Since, the common overlap begins at column $L=q$, the corresponding left boundary inside this shifted copy is $L-p=q-p=s$. Therefore, $V(p,i)=\lprev^{L-p}_P[i-p]=\lprev^{s}_P[s+t]$. For the shifted copy beginning at $q$, the checkpoint column $i=q+t$ corresponds to position $i-q=t$ inside the copy of $P$. The left boundary inside this shifted copy is $L-q=q-q=0$. Therefore, $V(q,i)=\lprev^{L-q}_P[i-q]=\lprev^{0}_P[t]$. Since $t$ is the parameterized witness for shift $s$, we have $\prev(P[0 ~\ldots~  m-s-1])[t] \neq \prev(P[s ~\ldots~  m-1])[t]$. Hence, $V(p,i) \neq V(q,i)$.
\end{proof}

For a checkpoint column $i$, we group the surviving shifts according to their checkpoint value $G_v = \{\gamma \in A_r: V(\gamma,i)=v\}$. Since \(V(p,i) \neq V(q,i)\), at least two groups are non-empty.
\begin{lemma}
\label{lem:random-chkpt-halve}
At round $r$, suppose the algorithm chooses $c$ uniformly at random from \{V(p,i),V(q,i)\}. Then, with probability at least 1/2, $|A_{r+1}| \leq \frac{|A_r|}{2}$.
\end{lemma}
\begin{proof}
    By lemma~\ref{lem:witness-separates}, the two values $V(p,i)$ and $V(q,i)$ are distinct. Therefore, the two groups $G_{(V(p,i))}$ and $G_{V(q,i)}$ are distinct and disjoint subsets of $A_r$. Hence, at least one of the groups has size at most $|A_r|/2$. The algorithm chooses one of the two values $V(p,i)$ and $V(q,i)$ uniformly at random. Therefore, with probability at least $1/2$, it chooses a value $c$ such that $|G_c| \leq \frac{|A_r|}{2}$. Since, next surviving set is exactly $A_{r+1}=G_c$, we obtain $|A_{r+1}| \leq \frac{|A_r|}{2}$ with probability at least $1/2$.
\end{proof}

\begin{definition}[Parameterized deterministic sample]
A parameterized deterministic
sample for a p-aperiodic pattern \(P[0 ~\ldots~  m-1]\) is a pair $(k; C_0,C_1,\ldots,C_{\ell-1}),$
where \(k < m/2\) is the unique surviving shift, and each checkpoint \(C_r\)
has the form $C_r = (L_r,i_r,c_r)$, where $L_r$ is the left boundary of the overlap in round $r$, $i_r$ is the global witness checkpoint column and $c_r$ is the checkpoint value. 
\end{definition} 
Each checkpoint \(C_r = (L_r, i_r,c_r)\) imposes a test on a shift $\gamma$. A shift \(\gamma\) satisfies checkpoint \(C_r\) if ${\lprev}^{\,L_r-\gamma}_{P}[i_r-\gamma] = c_r.$  The deterministic sample is constructed so that exactly one shift \(k\) satisfies all checkpoint tests:
\[
\bigcap_{r=0}^{\ell-1}
\left\{
\gamma :
\lprev_P^{L_r-\gamma}[i_r-\gamma] = c_r
\right\}
=
\{k\}.
\]
In the algorithm, we represent the surviving set $A_r$ implicitly using a predicate $\Alive_r$, where $A_r=\{\gamma:\Alive_r(\gamma)=1\}.$ \\

\begin{algorithm}[H]
\caption{Quantum-Friendly Parameterized Deterministic Sampling}
\label{alg:pdet-sample}

\KwIn{A p-aperiodic pattern $P[0 ~\ldots~  m-1]$}
\KwOut{A parameterized deterministic sample $\mathcal{D}$ with high probability}

Precomputed $\prev_T[\cdot]$ and $\prev_P[\cdot]$\;

$\mathcal{D}\gets \emptyset$\;
Choose a sufficiently large constant $\alpha$ and set $R=\alpha\log m$

Define the initial survival predicate $
\Alive_0(\gamma)=1
\quad\Longleftrightarrow\quad
0\leq \gamma < \lfloor m/2\rfloor .
$

\For{$r=0,1,\ldots,R-1$}{
\If{there is at most one $\gamma$ such that $\Alive_r(\gamma)=1$}{break\;}

    Let $p \gets \min\{\gamma:\Alive_r(\gamma)=1\}$
    and
    $q \gets \max\{\gamma:\Alive_r(\gamma)=1\}$.

    $s\gets q-p$\;

    Find a parameterized witness position $t$ for shift $s$, such that, $
    \prev(P[0 ~\ldots~  m-s-1])[t]
    \neq
    \prev(P[s ~\ldots~  m-1])[t]$.

    $L\gets q$\; $i\gets q+t$\;

    Compute $
    v_p\gets \lprev^{L-p}_P[i-p]
    $
    and
    $
    v_q\gets \lprev^{L-q}_P[i-q].
    $

    Choose $c$ uniformly at random from $\{v_p,v_q\}$\;

    Add checkpoint $C_r=(L,i,c)$
    to $\mathcal{D}$\;

    Define the next survival predicate implicitly by adding the checkpoint $c$, $
    \Alive_{r+1}(\gamma)
    =
    \Alive_r(\gamma)
    \wedge
    \left[
    \lprev^{L-\gamma}_P[i-\gamma]=c
    \right].
    $
}

Let $\ell$ be the number of checkpoints added.
If there is a unique shift $k$ such that $\Alive_{\ell}(k)=1$, then return $\mathcal{D}=(k;C_0,C_1,\ldots,C_{\ell-1})$, otherwise return failure.

\end{algorithm}
For any round $r$, a shift $\gamma$ is alive if and only if it satisfies all the checkpoints selected so far, i.e. $\Alive_r(\gamma)=1 \Longleftrightarrow 0 \leq \gamma < \lfloor m/2 \rfloor$ and $ \forall j<r, \; \lprev_P^{L_j-\gamma}[i_j-\gamma]=c_j,$ where $C_j=(L_j,i_j,c_j)$.

\begin{theorem}[Correctness of Algorithm~\ref{alg:pdet-sample}]
\label{thm:alg1-correctness}
Let $P[0 ~\ldots~  m-1]$ be p-aperiodic. Then Algorithm~\ref{alg:pdet-sample} returns a valid parameterized deterministic sample of size $O(\log m)$ with high probability.
\end{theorem}
\begin{proof}
Let $A_r$ be the set of shifts that are alive after $r$ rounds. Initially, $|A_0| 
\leq m/2$. By Lemma~\ref{lem:random-chkpt-halve}, each round reduces the number of alive shifts by a factor or two with probability at least $1/2$. So, after $\lceil \log_2 m\rceil$ successful rounds, the alive set has size at most one. 
The algorithm performs $R=\alpha \log m$ rounds. For a sufficiently large constant $\alpha$, a standard Chernoff bound implies that at least $\lceil \log_2 m\rceil$ successful rounds occur with probability at least $1-m^{-\Omega(1)}$. Hence, with high probability, the final alive set contains a unique shift $k$.
By construction, a shift remains alive if and only if it satisfies all selected checkpoints. Therefore, the returned tuple $D=(k;C_0,C_1,\ldots,C_{\ell-1})$ is a valid parameterized deterministic sample. Since the algorithm performs only $R=O(\log m)$ rounds, the sample size is $\ell=O(\log m)$.
\end{proof}

\begin{lemma}
\label{True-occ-passes-sample}
Let $D=(k;C_0,C_1,\ldots,C_{\ell-1})$ be a valid parameterized deterministic sample for $P[0 ~\ldots~  m-1]$. If $P$ occurs at alignment $a$ in $T$ then $O_{\mathrm{sample}}(a)=1.$   
\end{lemma}
\begin{proof}
Let $C_r=(L_r,i_r,c_r)$ be any checkpoint. Since, $D$ is a valid sample, the surviving shift $k$ satisfies $\lprev^{L_r-k}_P[i_r-k]=c_r$. Since $P\approx_p T[a ~\ldots~  a+m-1]$, the corresponding local previous-encoding values in the text and pattern are equal. So, $\lprev^{a+L_r-k}_T[a+i_r-k]=\lprev^{L_r-k}_P[i_r-k]=c_r.$ This holds for every checkpoint $C_r$. Therefore, $O_{\mathrm{sample}}(a)=1.$
\end{proof}

\begin{lemma}
\label{pDS-ricochet-property}
Let $D=(k;C_0,C_1,\ldots,C_{\ell-1})$ be a valid parameterized deterministic sample for $P[0 ~\ldots~  m-1]$. Suppose an alignment $x$ satisfies the sample test, i.e., $O_{\mathrm{sample}}(x)=1$. Let $y$ be another alignment, and define $\gamma = k+y-x$. If $0\leq \gamma < \lfloor m/2\rfloor$ and $P$ occurs at alignment $y$, then $y=x$.
\end{lemma}
\begin{proof}
Let $C_r=(L_r,i_r,c_r)$ be any checkpoint. Since $O_{\mathrm{sample}}(x)=1$, we have 
\[\lprev^{x+L_r-k}_T[x+i_r-k]=c_r\]. Using $\gamma=k+y-x$, this is the same as $\lprev^{y+L_r-\gamma}_T[y+i_r-\gamma]=c_r$. Since, $P$ occurs at alignment $y$, the corresponding local previous-encoding values in the text and pattern are equal. Therefore, $\lprev^{L_r-\gamma}_P[i_r-\gamma]=c_r$. Thus $\gamma$ satisfies checkpoint $C_r$. Since, this holds for every checkpoint, $\gamma$ satisfies all checkpoints of $D$. But $D$ is valid, so the only shift satisfying all checkpoints is $k$. Hence, $\gamma=k$. Since, $\gamma=k+y-x$, we get $y=x$.
\end{proof}

\subsection{Algorithm for Parameterized Pattern Matching}
In this section, we give the parameterized pattern matching algorithm using the parameterized deterministic sample constructed in Algorithm 1.

We partition the text into blocks of length $B=\lfloor\frac{m}{4}\rfloor$. For a block index $b$, define the set of valid alignments in block $b$ by $I_b=\left\{a:bB\leq a<\min\{(b+1)B,n-m+1\} \right\}.$ Thus, every alignment in $I_b$ is a valid starting position for an occurrence of $P$ in $T$. Algorithm~2 shows the full quantum-friendly algorithm for parameterized pattern matching.\\

\begin{lemma}
\label{lem:blk-oracle-corretness}
For every block $b$, $O_{\mathrm{block}}(b)=1$ if and only if block $b$ contains a parameterized occurrence of $P$.
\end{lemma}
\begin{proof}
Let $I_b$ be a set of valid alignments in block $b$. First, suppose $O_{\mathrm{block}}(b)=1$. Then, by definition of the block oracle, the inner search found an alignment $a \in I_b$ satisfying $O_{\mathrm{sample}}(a)=1$ and the block oracle then accepted after applying full verification to $a$. Hence, $O_{\mathrm{verify}}(a)=1$. By definition of full verification, this means $\prev(P)=\prev(T[a ~\ldots~  a+m-1])$. Therefore, $P\approx_p T[a ~\ldots~  a+m-1]$. Thus, block $b$ contains a parameterized occurrence of $P$.

Conversely, suppose that block $b$ contains a true parameterized occurrence at alignment $a$ in $I_b$. By lemma~\ref{True-occ-passes-sample}, this alignment passes the test $O_{\mathrm{sample}}(a)=1$. Therefore, the inner search over $I_b$ finds some sample-passing alignment $a' \in I_b$. Since $a$ and $a'$ are in the same block, we have $|a-a'| < B$. By choice $B=\lfloor m/4 \rfloor$, we have  $|a-a'| < m/2$. Hence, by lemma~\ref{pDS-ricochet-property}, the sample-passed alignment found in this block must be the true occurrence itself. Therefore, $a'=a$. Since $a$ is a true occurrence, full verification accepts $a'$. So, $O_{\mathrm{block}}(b)=1$. Therefore, $O_{\mathrm{block}}(b)=1$ if and only if block $b$ contains a parameterized occurrence of $P$.
\end{proof}

\setcounter{AlgoLine}{0}
\begin{algorithm}[H]
\label{alg2}
\caption{Parameterized Pattern Matching for p-Aperiodic Patterns}
\label{alg:quantum-pmatch}

\KwIn{Text $T[0 ~\ldots~  n-1]$, p-aperiodic pattern $P[0 ~\ldots~  m-1]$, and deterministic sample
$\mathcal{D}=(k;C_0,C_1,\ldots,C_{\ell-1})$}
\KwOut{A position $a$ such that $P\approx_p T[a ~\ldots~  a+m-1]$, or $\mathsf{NoMatch}$}

\If{$m>n$}{
    \Return $\mathsf{NoMatch}$\;
}

Set block length $B\gets \left\lfloor \frac{m}{4}\right\rfloor$, number of blocks $
N_B\gets \left\lceil \frac{n-m+1}{B}\right\rceil$\;

For each block index $b=0,1,\ldots,N_B-1$, define $
    I_b
    =
    \left\{
    a:
    bB\le a<\min\{(b+1)B,n-m+1\}
    \right\}.$

Define the block oracle \(\Oblock(b)\) as follows\;
\Indp
Use Grover search over \(a\in I_b\) to find an alignment satisfying \(\Osample(a)=1\)\;
\If{no such alignment exists}{
    \Return \(0\)\;
}
Let \(a_b\) be the sample-passing alignment found\;
\If{\(\Overify(a_b)=1\)}{
    \Return \(1\)\;
}
\Else{
    \Return \(0\)\;
}
\Indm

Use Grover search over block indices \(b=0,1,\ldots,N_B-1\) with oracle \(\Oblock(b)\)\;

\If{no accepting block is found}{
    \Return \(\mathsf{NoMatch}\)\;
}

Let \(b'\) be an accepting block\;
Use Grover search over \(a\in I_{b'}\) to find an alignment satisfying \(\Osample(a)=1\)\;
Let \(a'\) be the sample-passing alignment found\;

\If{$\Overify(a')=1$}{
    \Return $a'$\;
}
\Else{
    \Return $\mathsf{NoMatch}$\;
}
\end{algorithm}

\begin{theorem}[Correctness of Algorithm~\ref{alg:quantum-pmatch}]
\label{thm:alg2-correctness}
Algorithm~\ref{alg:quantum-pmatch} returns a valid parameterized occurrence if one exists, and returns $\mathsf{NoMatch}$ otherwise, upto the bounded error probability of the quantum search procedure.    
\end{theorem}

\begin{proof}
The algorithm searches over block indices using the block oracle $O_{\mathrm{block}}$. By lemma~\ref{lem:blk-oracle-corretness}, a block is accepting exactly when it contains a parameterized occurrence of $P$. If a parameterized occurrence exists, then it belongs to some block $b$. By lemma~\ref{lem:blk-oracle-corretness}, this block is accepting, and therefore the outer Grover search finds an accepting block with bounded error probability. After an accepting block is found, the algorithm again searches within that block for a sample-passing alignment and then applies full verification to the alignment found. If the algorithm returns an alignment $a$, then $O_{\mathrm{verify}}(a)=1$. Hence, $\prev(P)=\prev(T[a ~\ldots~  a+m-1])$ and so $P \approx_p T[a ~\ldots~  a+m-1]$. Thus, every returned alignment is a valid parameterized occurrence. 
If no parameterized occurrence exists, then by lemma~\ref{lem:blk-oracle-corretness}, no block is accepting. Hence, the outer Grover search finds no accepting block, and the algorithm returns $\mathsf{NoMatch}$. Therefore, Algorithm~\ref{alg:quantum-pmatch} is correct, up to the bounded error probability of the quantum search procedures.
\end{proof}

\subsection{Complexity}
\label{sec:Complexity}
Recall that throughout the analysis, we assume QRAM access to the input text pattern, and precomputed arrays $\prev[\cdot]$ of the input text and pattern. Under this assumption, the required local previous-encoding values can be accessed in constant time. The logarithmic factors arising from bounded-error quantum search and error-reduction are hidden in $\widetilde{O}(\cdot)$ notation.\\
\paragraph{Sample Construction Complexity.} Let $R=\alpha\log m$ be the number of sampling rounds. In each round, the algorithm maintains the set of surviving shifts implicitly by the predicate $\Alive_r(\gamma)$. To find the extreme surviving shifts, the algorithm computes $p=\min\{\gamma:\Alive_r(\gamma)=1\}$ and $q=\max\{\gamma:\Alive_r(\gamma)=1\}.$ These can be found using the quantum minimum and maximum finding algorithm over the range $0 \leq \gamma < \lfloor \frac{m}{2} \rfloor$. For a fixed shift $\gamma$, evaluating $\Alive_r(\gamma)$ requires checking the $r$ previously chosen checkpoints. Since $r \leq R = O(\log m)$, this costs $O(r)$ local previous-encoding queries if the checkpoints are checked directly. Alternatively, by Grover search over the checkpoints, one can check whether any checkpoint is violated in $O(\sqrt{r})$ queries. Therefore, the cost of finding $p$ and $q$ in round $r$ is $O(\sqrt{m}\sqrt{r})$ local previous-encoding queries, up to logarithmic factors from the bounded-error search. Since $r=O(\log m)$, this is $O(\sqrt{m\log m})$ = $\widetilde{O}(\sqrt m)$. After finding $p$ and $q$ , the algorithm sets $s=q-p$ and searches for a parameterized witness position $t$ that satisfies $\prev(P[0 ~\ldots~  m-s-1])[t] \neq \prev(P[s ~\ldots~  m-1])[t]$. This witness search covers at most $m$ positions, and each witness test can be evaluated using constant-time local previous-encoding queries, assuming that the required previous-occurrence arrays are available. Hence, the witness search costs $O(\sqrt{m})$ queries. Thus, one round costs $\widetilde{O}(\sqrt m)$. Since there are rounds $R=O(\log m)$, the total cost of constructing the deterministic sample is $\widetilde{O}(\sqrt m)$. The sample size is at most $O(\log m)$ with high probability. By choosing the constant $\alpha$ in $R=\alpha \log m$ large enough, the probability that fewer than $\log m$ successful halving rounds occur is made polynomially small. Therefore, assuming $O(1)$-time access to local previous-encoding, Algorithm~1 constructs a parameterized deterministic sample of size $O(\log m)$ with high probability in $\widetilde{O}(\sqrt m)$ time.\\

\paragraph{Pattern Matching Query Complexity.}Let $\ell=O(\log m)$ be the number of checkpoints in the deterministic sample. To check whether a fixed alignment $a$ satisfies the sample requires testing all $l$ checkpoints. Using Grover search over the checkpoints to search for a violated checkpoint costs $O(\sqrt{\ell}) = O(\sqrt{\log m})$ queries to the text local-prev oracle. Within one block, the number of possible alignments is $|I_b| \leq B = O(m)$. Therefore, finding a sample candidate in a block by Grover search costs $O(\sqrt{B}\cdot \sqrt{\ell}) = O(\sqrt{m}\sqrt{\log m})=\widetilde{O}(\sqrt m)$. Verifying a candidate alignment requires searching for a mismatch over m positions. Thus, verification costs $O(\sqrt{m})$ and an evaluation of block oracle costs $\widetilde{O}(\sqrt m)$. The number of blocks is $N_B = O(\frac{n}{m})$. The Grover search over the block oracle therefore costs $O\left( \sqrt{\frac{n}{m}} \cdot \widetilde{O}(\sqrt m) \right) = \widetilde{O}(\sqrt n).$ After an accepting block is found, the algorithm repeats the inner sample and performs one final verification. This adds $\widetilde{O}(\sqrt m)$ queries, which is included in the total $\widetilde{O}(\sqrt n+\sqrt m)$ bound.

\begin{theorem}[Total query complexity for p-aperiodic pattern matching]
Assuming QRAM access to the inputs and to the precomputed previous-encoding arrays of the input text $T[0 ~\ldots~  n-1]$ and a p-aperiodic pattern $P[0 ~\ldots~  m-1]$, algorithms~\ref{alg:pdet-sample} and ~\ref{alg:quantum-pmatch} solve parameterized pattern matching problem with bounded error in $\widetilde{O}(\sqrt m+\sqrt n)$ queries.   
\end{theorem}
\begin{proof}
The sample construction costs $\widetilde{O}(\sqrt m)$ queries and the matching phase costs $\widetilde{O}(\sqrt n)$ queries by the complexity analyses in section ~\ref{sec:Complexity}. Therefore, the total complexity is $\widetilde{O}(\sqrt m)+\widetilde{O}(\sqrt n)=\widetilde{O}(\sqrt m+\sqrt n)$. The correctness follows from Theorem~\ref{thm:alg1-correctness} and Theorem~\ref{thm:alg2-correctness}. 
\end{proof}
\section{Conclusions and Future Work}

Our result provide a step towards understanding how quantum algorithms can be used for parameterized pattern matching. We obtain query complexity of $\widetilde{O}(\sqrt{n}+\sqrt{m})$ for p-aperiodic pattern matching, assuming that the previous-encoding of text and pattern are available as an Oracle.

There several interesting directions for further investigation. First, our current algorithm is restricted to p-aperiodic patterns. In classical exact string matching, Vishkin’s deterministic sampling~\cite{Vishkin1991} with quantum adaptations~\cite{RameshVinay2003, wang2024quantum} uses a small sample of size $O(\log m)$ to eliminate all but one candidate shifts when the pattern is aperiodic, while periodic patterns can be handled using classical periodicity lemmas. In parameterized setting, this approach is more complicated because parameterized strings do not always follow classical periodicity properties and may exhibit multiple pseudo-periods~\cite{Amir1994,ApostolicoG08}. Furthermore, candidate shifts can show irregular behavior before settling into a periodic progression~\cite{JalseniusPS13}, making periodic parameterized pattern matching more challenging. Extending it to p-periodic patterns would be an important step towards handling general parameterized pattern matching. Second, it is interesting to develop a quantum algorithm that works directly on the input text and pattern, without relying on precomputed previous encodings. This would make the approach more applicable to general quantum oracle models. However, removing the precomputed previous-encoding assumption introduces an additional difficulty. Parameterized matching over an unrestricted alphabet contains the element distinctness problem~\cite{AaronsonShi2004} which requires $\Omega(m^{2/3})$ quantum queries in the worst case. Exploring how to overcome these challenges remains an interesting direction for future research.

\bibliographystyle{abbrv}
\bibliography{reference}

\appendix
\section{Oracle Definitions}
\label{app:oracles}

\subsection{Oracles}
In this section, we describe the oracles used by the quantum algorithm for parameterized deterministic sampling and matching. The main difference from exact matching is that the value observed at a position is not simply the input character. Instead, the relevant value is the local previous-encoding value, which depends on the current sub-string boundary. We define oracles for previous-encoding, alive-shift testing, witness finding, sample checking and full verification.\\

\begin{enumerate}

\item \textbf{Text Local-Prev Oracle.} For a text boundary $b$ and a text position $y$, where $0 \leq b \leq y < n $, the text local-previous-encoding oracle returns 
\[
\Otlprev(b,y)= \lprev^b_T[y].
\]
Using the precomputed previous-encoding array, this value is given by
\[\Otlprev(b,y)=
\begin{cases}
0, & \prev_T[y] > y-b,\\[2mm]
\prev_T[y], & \prev_T[y] \le y-b.
\end{cases}
\]
This oracle is used during sample checking and full verification to obtain the local previous-encoding value of a text position with respect to a specified alignment boundary.\\

 \item \textbf{Pattern Local-Prev Oracle.}
Given a pattern position $i$ and a left boundary $b$, where $0 \leq b \leq i < m$, the pattern local-previous-encoding oralce returns $\lprev^{b}_{P}[i]$. Using the precomputed local previous-encoding array, this value is given by
\[
\Oplprev(b,i)=
\begin{cases}
0, & \prev_P[i]>i-b,\\[2mm]
\prev_P[i], & \prev_P[i]\leq i-b.
\end{cases}
\]
This oracle is used during deterministic sampling. A checkpoint
$C_r=(L_r,i_r,c_r)$ is satisfied by a shift $\gamma$ if and only if $\Oplprev(L_r-\gamma,i_r-\gamma)=c_r.$

\item \textbf{Witness Oracle.}
Let $1\leq s\leq \lfloor m/2 \rfloor$ be a shift. A position $t$, where $0\leq t<m-s$, is a parameterized witness for $s$ if $ \prev(P[0 ~\ldots~  m-s-1])[t]\neq \prev(P[s ~\ldots~  m-1])[t].$ The witness oracle is defined as
\[
\Owit(s,t)=
\begin{cases}
1, & \text{if }
\prev(P[0 ~\ldots~  m-s-1])[t]\neq \prev(P[s ~\ldots~  m-1])[t],\\[2mm]
0, & \text{otherwise.}
\end{cases}
\]
In the sampling algorithm, this oracle is invoked with $s=q-p$ where $p$ and $q$ are the minimum and maximum alive shifts in the current round. Since $P$ is p-aperiodic, every shift $1\leq s\leq \lfloor m/2\rfloor$ has at least one witness. Therefore, for any such shift $s$, Grover search can be used with $\Owit$ to find a witness position $t$.
Equivalently, we may view the witness oracle as comparing the two shifted
copies determined by $p$ and $q$. Let $s=q-p$. Then the same test is $
\Owit(p,q,t)=1$ if and only if $\prev(P[0 ~\ldots~  m-s-1])[t]\neq \prev(P[s ~\ldots~  m-1])[t].$
Thus, the witness oracle detects a mismatch in local previous-encoding
between the two overlapping copies. This is a structural witness, not a raw character mismatch.\\

\item \textbf{Sample-checking Oracle.}
When a sample is checked against a text alignment $a$, the corresponding text boundary is $a+L_r-k$ and the corresponding text position is $a+i_r-k$. Thus, alignment $a$ satisfies checkpoint $C_r$ if $\Otlprev(a+L_r-k,\; a+i_r-k)=c_r.$
The sample-checking oracle is defined by
\[
\Osample(a)=
\begin{cases}
1, & \Otlprev(a+L_r-k,\;a+i_r-k)=c_r,
      \text{ for every } r=0,1,\ldots,\ell-1,\\[2mm]
0, & \text{otherwise}.
\end{cases}
\]

\item \textbf{Alive-shift Oracle.}
During the sampling process, the surviving set of shifts is represented by a predicate. Let $ \mathcal{D}_r=(C_0,C_1,\ldots,C_{r-1})$ be the checkpoints selected during the first $r$ rounds, where $C_j=(L_j,i_j,c_j).$
We define a parameter $Q$ that restricts the current shift range. Initially, $ Q=\lfloor m/2\rfloor-1.$
The alive-shift oracle is defined as
\[
\Oalive(r,\gamma,Q)=
\begin{cases}
1, & 0\leq \gamma\leq Q \text{ and }
      \Oplprev(L_j-\gamma,i_j-\gamma)=c_j
      \text{ for every } j<r,\\[2mm]
0, & \text{otherwise}.
\end{cases}
\]

Equivalently, $ \Oalive(r,\gamma,Q)=1$ exactly when shift $\gamma$ has not been eliminated by the first $r$ checkpoints. This oracle allows the algorithm to search over surviving shifts without
explicitly storing the entire set of alive shifts. The sampling
algorithm uses it to find $p=\min\{\gamma:\Oalive(r,\gamma,Q)=1\}$ and $ q=\max\{\gamma:\Oalive(r,\gamma,Q)=1\}.$\\

 \item \textbf{Parameterized Mismatch Oracle.}
The sample-checking oracle filters the candidate alignments in the text. To verify that a candidate alignment is a full parameterized occurrence of the pattern, we search for a mismatch in local previous-encoding. For an alignment $a$ and a relative position $x$, where $0 \leq x < m$, define the mismatch oracle as
\[
\Omis(a,x)=
\begin{cases}
1, & \text{if } \prev_P[x]\neq \OlprevT(a,x)\\[2mm]
0, & \text{otherwise.}
\end{cases}
\]

 \item \textbf{Verification Oracle.} The verification oracle accepts an alignment if no mismatch exists. Formally, 
\[
\Overify(a)=
\begin{cases}
1, & \Omis(a,x)=0 \text{ for every } x=0,1,\ldots,m-1,\\[2mm]
0, & \text{otherwise}.
\end{cases}
\]
Equivalently, $ \Overify(a)=1 \Longleftrightarrow \prev_P[x]=\OlprevT(a,x) \qquad \text{for all } x\in[0 ~\ldots~  m-1].$\\

 \item \textbf{Block Oracle.}
The quantum matching algorithm partitions the text into blocks of length $B=m/4$. Let $I_b$ denote the set of valid starting positions inside block $b$. The block oracle determines whether the block $b$ contains a valid parameterized occurrence. Formally, the block oracle is defined as 
\[
\Oblock(b)=
\begin{cases}
1, & \text{there exists an alignment } a\in I_b
      \text{ such that } \Osample(a)=1
      \text{ and } \Overify(a)=1,\\[2mm]
0, & \text{otherwise}.
\end{cases}
\]
Equivalently, $\Oblock(b)$ accepts a block when the block contains at least one alignment that passes the deterministic-sample filter and is then verified as a true parameterized occurrence.
\end{enumerate}
\end{document}